# The Price of Using Students

Dag I.K. Sjøberg and Gunnar R. Bergersen

**Abstract**—In a recent article, Falessi et al. (2017) call for a deeper understanding of the pros and cons of using students and professionals in experiments. The authors state: "we have observed too many times that our papers were rejected because we used students as subjects." Good experiments with students are certainly a valuable asset in the body of research in software engineering. Papers should thus not be rejected solely on the ground that the subjects are students. However, the distribution in skill is different for students and professionals. Since previous studies have shown that skill may have a moderating effect on the treatment of participants, we are concerned that studies involving developers with only low to medium skill (i.e., students) may result in wrong inferences about which technology, method or tool is better in the software industry. We therefore provide suggestions for how experiments with students can be improved and also comment on some of the alleged drawbacks of using professionals that Falessi et al. point out.

———————————— ◆ ————————————

## 1 INTRODUCTION

While we know that the level of skill among subpopulations of students varies substantially, less attention has been paid to the variation in skill among subpopulations of professionals, which is also substantial. For example, we previously found that the benefit of using pair programming was positive for juniors, whereas it was negative for senior developers (Arisholm et al. 2007). In another experiment, only the best performing developers benefitted from a "proper" object-oriented control style, whereas the others did not (Arisholm and Sjøberg 2004). Also, in a study by Krein et al. (2016), the purported benefit of the investigated design patterns was mostly negative for the students but mostly positive for the most experienced and knowledgeable professionals. Such experiments show that one technology (method, technique, tool, etc.) may be better for developers at one skill level, while another technology may be better for developers at another skill level, as illustrated in Figure 1. This "reversal effect" (Sjøberg et al. 2016) needs to be taken into account when designing and analyzing experiments.

Another reason for the apparently contradictory results in the literature on the differences between students and professionals is that the skill distribution of various subpopulations is rarely taken into account. Figure 2 illustrates the distributions of five categories of developers. The graphs representing junior, intermediate and senior professionals are based on aggregated data from three experiments (n = 262) lasting one or two days (Arisholm and Sjøberg 2004, Arisholm et al. 2007, Bergersen et al. 2014).

The exact form of the curves of the undergraduate and graduate student populations are hypothesized because we do not have comprehensive data from multiple experiments that include both students and professionals, but the mean skill of students is placed to the left of the mean skill of professionals (i.e., less skilled) based on the well established *theoretical consideration* that people improve their skills through practice, cf. the law of practice (Newell and Rosenbloom 1981). Note that this theoretical consideration is based on hundreds of empirical studies in various disciplines. If one thinks in terms of *cohorts*, developers undergo transitions from undergraduate to graduate students and then to junior, intermediate, and senior professionals, working with increasingly greater skill throughout their education and their professional career.

In practice, however, many developers do not transition through all the categories. For example, some undergraduates start working as juniors after their BSc degree; some companies do not have an "intermediate" professional developer category; some graduates may advance directly into the intermediate category; and so on. In addition, while published experiments show that the mean skill of undergraduates is generally lower than that of professionals, the mean skill of graduate students might be higher than that of junior professionals, depending on the concrete topic of investigation and domain of skill. In any case, the overall distribution of students will be to the left of the overall distribution of professionals (see the shaded areas). Consequently, one should be cautious about generalizing when the sample is drawn from a population (students) other than that which one aims to generalize to (professionals).

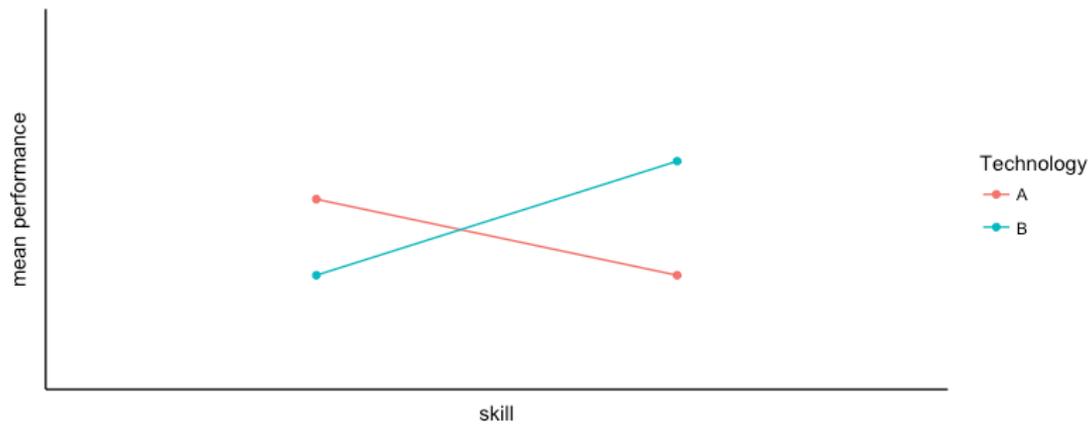

**Fig. 1** Which technology is best depends on skill level

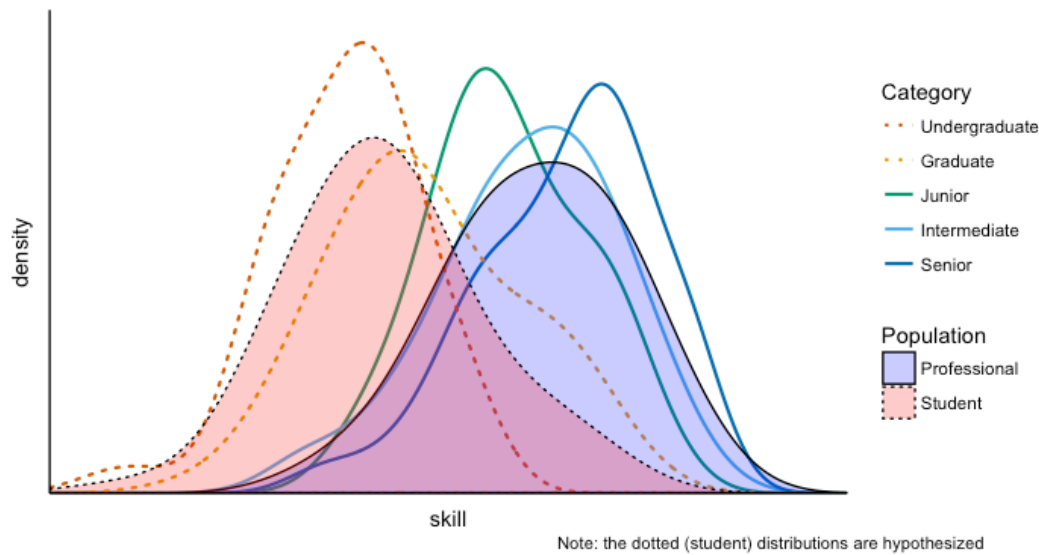

**Fig. 2** Distribution of skill in different populations

For any specific and well-defined skill, individual differences can be substantial. Moreover, different skills will be distributed differently, and which skill is relevant in a given experiment will depend on the experimental treatment. For example, if the technology being evaluated in an experiment recently had been taught to a group of students, they may be *more* skilled in that technology than a group of professionals who had not recently used that technology. Such a case supports the argument of Falessi et al. that it is not always better to experiment with professionals.

The sample size of most studies in software engineering is small and subjects are sampled by convenience. The samples are therefore rarely representative of the underlying student or professional populations. Highly skilled students and less skilled professionals will be sampled on occasion. A study may then easily show that students perform better than or similar to professionals. Similar results between students and professionals were obtained by, for example, Höst et al. (2000), which might have been surprising at the time but may be expected given non-random sampling from overlapping distributions (Figure 2). Moreover, the study contrasted (subjective) judgments of students versus professionals and not (objective) performance-based measures where skill is more central. Thus, we caution against using a few (outlier) studies to justify that students are good proxies for professionals in general or in a given study.

Falessi et al. state that "many more comparative studies are needed before we obtain an answer on whether students are good proxies of professionals in software engineering experiments." Since students and professionals have different characteristics, one can hardly ever claim that students are good proxies for professionals *in general*, but certain student samples may be good proxies for certain subsets of professionals. Careful empirical investigation is indeed needed to detect those cases.

## 2 ALLEGED DRAWBACKS OF PROFESSIONALS

Falessi et al. conducted a survey among empirical researchers and reported that the respondents in general disagreed with Falessi et al. about the drawbacks of professionals. We also have several objections to the drawbacks stated by Falessi et al. First, they claim that paying professionals for participating in experiments is a "strong threat to validity." The Merriam-Webster dictionary defines "professional" work as being "engaged in by persons receiving financial return." Professionals are thus by definition paid for their work, whether for work done in experimental or non-experimental settings. The money is not additional payment to individuals to work extra in their free time, but rather compensates the organization for those who participate in a study during ordinary (paid) work hours. A sample may easily be biased if only those who are willing to spend their spare time are included in the sample because such volunteers may possess characteristics that affect the experimental outcome that are different from the characteristics of the typical paid industrial developer. In the same vein, students may participate in experiments as part of mandatory coursework or may be rewarded by extra course credits. In our view, compensation for participation in experiments motivates developers in a way similar to that of ordinary work, thus reducing the likelihood of confounding effects of differences in motivation between the available sample and the target population.

Second, Falessi et al. state that there is a tradeoff between higher *internal* validity when using students versus higher *external* validity when using professionals. Clearly, there are fewer concerns about external validity in experiments with professionals, but the moderating effect that skill may have on the treatment of an experiment (see Section 1) is a particular threat to the internal validity of student-based experiments. Moreover, it is certainly possible to have better internal validity in a well-designed student study than in a poorly designed professional study, but that does not mean that student experiments have better internal validity per se. To support the claim that internal validity is lower when using professionals, Falessi et al. refer to a book on research methods in health sciences, which states that "the relationship between internal and external validity is inverse" (Berg and Latin 2003, p. 213). According to this premise, any student experiment would have higher internal validity than any experiment with professionals, which is clearly not the case. Software engineering is not a typical lab science; the same kind of treatment and experimental control can be applied to professionals as well as to students. The level of internal validity may be the same in a student experiment as in an experiment with professionals, but the latter may have higher external validity.

Third, another drawback to using professionals that Falessi et al. claim is that sample sizes are small and that professionals tend not to show up for experiments. Without any budget for recruiting subjects, it is certainly easier to obtain larger sample sizes with students. It is easier to obtain many mice for a medical experiment, but medical researchers do not consider using humans to be a drawback. Moreover, if one follows a well-designed procedure for recruiting professionals, including contracted payment with the organizations of the professionals and conduct the experiments in regular work hours (Sjøberg et al. 2007), they will show up. In our own experiments, the no-shows are negligible and comparable to the percentage of individuals that would be on sick leave for any given day.

Fourth, Falessi et al. claim that professionals are less committed than students to finish tasks on time because students "are used to an examination culture." Finishing tasks on time and keeping deadlines are certainly important in work life, and this claim is in stark contrast to what was the case in our experiments with hundreds of professionals.

## 3 CHARACTERIZING SUBJECTS

We agree with Falessi et al. that our research field must characterize subjects beyond whether they are students or professionals. Based on a suggestion from one of the respondents in their study, Falessi et al. propose that subjects should be interviewed and qualitatively described using a characterization of real, relevant and recent experience ($R^3$). However, they define real experience as "experience judged relevant by the researcher in dialog with each subject," indicating that there is no difference between real and relevant experience in their definition. Nevertheless, there is empirical support for using recent experience (e.g., Sigmund et al. 2014), in particular when there is a good match in specificity between the experience predictor and the criterion being predicted (Bergersen et al. 2014). For example, if a study requires Java programming, "*Java* programming experience" will be a better predictor than "*general* programming experience."

Additionally, Falessi et al. suggest replacing a student-professional dichotomy with an experience trichotomy (0–2, 3–5, 5+ years of experience). Discretizing a continuous variable such as experience reduces statistical power and is, thus, a threat to statistical conclusion validity (Shadish et al. 2002). Furthermore, using years of experience may work well for professionals but not for students, who mostly have none. Because variables with little variance or with ceiling/floor effects reduce statistical power (Shadish et al. 2002), the experience of students should be measured in months.

Even though experience plays an important role in both theories of skill (Fitts and Posner 1967) and expertise (Ericsson and Charness 1994), it is knowledge, skill and motivation that cause subjects to perform well (Campbell et al. 1993). Thus, experience is a proxy for skill, which together with a long list of other variables can be used to predict how well a programmer will perform in an experiment (see Bergersen et al. 2014). Unfortunately, although experience has a medium positive correlation with performance (around 0.30 across our own studies), it is neither the best nor the only proxy one can use (one should consider, e.g., LOC and self-reported skills).

A better alternative to using proxy variables is to develop and use pretests. They enable powerful designs such as matching, stratifying or blocking (Shadish et al. 2002). An example of the use of a small, single-task pretest can be found in (Arisholm and Sjøberg 2004). The main experiment consisted of a set of change tasks on a Java program that was provided in two variants of control style (the treatment of the experiment). Before starting on the experiment tasks, all participants performed the same change task on another Java program (with no variants). Individual differences in performance on this pretest task were used in the analysis of the experimental task to adjust for differences between the treatment groups. An example of the use of a more comprehensive pretest is described in (Bergersen and Sjøberg 2012).

Using posttests may also support the investigation of several potential threats to validity (for details, see Kampenes et al. 2009). For example, we have used pre- and posttests to investigate the effects of learning (practice) during an experiment, which may undermine the claim that subjects are already highly skilled (often called "experts") on the technology they are working on (Sheil 1981). In a sample of 65 professional developers, the level of performance of subjects was stable after a small warm-up period, thus demonstrating that the professionals did not learn anything during the experiment (Bergersen et al. 2014).

An alternative to developing one's own pre- and posttests is to use validated instruments that previously have been shown to highly correlate with the dependent variable in an experiment and where the skill level of different categories of professional developers are available. We have developed such an instrument to predict programming performance (Bergersen et al. 2014). Admittedly, there are few such instruments available, and they require much effort to develop.

## 4 CONCLUSION

Overall, we are concerned that the low proportion of professionals as participants reduces the impact of software engineering experiments on industry. It is difficult to influence development practices by arguing that a group of students benefitted from using a certain method or tool when it is unknown how these students differ in skill and motivation relative to professional programmers. Other things being equal, sampling from the same population that one aims to generalize to reduces threats to validity. In an earlier literature review, we found that only 9% of the subjects were professionals (Sjøberg et al. 2005). More recently, we examined the experiments published in the journals *IEEE Transactions on Software Engineering* and *ACM Transactions on Software Engineering and Methodology* in 2015 and 2016, and *Empirical Software Engineering* in 2016 and 2017. Among the total of 1752 subjects, only 139 were professionals, that is, 8%. Consequently, we urge the community to run more experiments with professionals.

Nevertheless, it may be impractical and expensive to obtain appropriate samples of professionals. Using students is then a good alternative, although the number and magnitude of potential threats to validity increase. Researchers should, therefore, carefully and cautiously report and discuss the limitations of student-based experiments. In particular, the moderating effect of skill level on the benefit of treatment should be addressed.